\documentclass[12pt,preprint]{aastex}
\usepackage{psfig,natbib}
\shortauthors{Bower et al.}
\shorttitle{Millimeter YSO Flare}
\begin{document}

\newcommand\degd{\ifmmode^{\circ}\!\!\!.\,\else$^{\circ}\!\!\!.\,$\fi}
\newcommand{\etal}{{\it et al.\ }}
\newcommand{\uv}{(u,v)}
\newcommand{\rdm}{{\rm\ rad\ m^{-2}}}
\newcommand{\msuny}{{\rm\ M_{\sun}\ y^{-1}}}
\newcommand{\mylesssim}{\stackrel{\scriptstyle <}{\scriptstyle \sim}}
\newcommand{\object}{GMR-A}

%\slugcomment{Accepted for publication in the Astrophysical Journal}

\title{A Giant Outburst at Millimeter Wavelengths in the Orion Nebula}

\author{Geoffrey C. Bower\altaffilmark{1}, 
Richard L. Plambeck\altaffilmark{1},
Alberto Bolatto\altaffilmark{1},
Nate McCrady\altaffilmark{1}, 
James R. Graham\altaffilmark{1},
Imke de Pater\altaffilmark{1}, 
Michael C. Liu\altaffilmark{2,3}, \&
Frederick K. Baganoff\altaffilmark{4}}

\altaffiltext{1}{Astronomy Department \& Radio Astronomy Laboratory, 
601 Campbell Hall,
University of California, Berkeley, CA 94720; gbower,plambeck,bolatto,nate,jrg,imke@astro.berkeley.edu} 
\altaffiltext{2}{Institute for Astronomy, University of Hawaii, 2680 Woodlawn Dr., Honolulu, HI 96822; mliu@ifa.hawaii.edu}
\altaffiltext{3}{Hubble Fellow}
\altaffiltext{4}{Center for Space
Research, Massachusetts Institute of Technology, Cambridge, MA 02139; fkb@space.mit.edu}

\begin{abstract}

BIMA observations of the Orion nebula discovered a giant flare 
from a young star  previously undetected at
millimeter wavelengths.   The star briefly became the brightest compact object in
the nebula at 86 GHz.  Its flux density increased by more than a factor
of 5 on a timescale of hours, to a peak of 160 mJy.  This is one of the
most luminous stellar radio flares ever observed.  Remarkably, the
Chandra X-ray observatory was in the midst of a deep integration of the
Orion nebula at the time of the BIMA discovery; the source's X-ray flux
increased by a factor of 10 approximately 2 days before the radio
detection.  Follow--up radio observations with the VLA and BIMA showed
that the source decayed on a timescale of days, then flared again
several times over the next 70 days, although never as brightly as
during the discovery.  Circular polarization was detected at 15, 22, and
43 GHz, indicating that the emission mechanism was cyclotron.  VLBA
observations 9 days after the initial flare yield a brightness
temperature $T_b>5\times10^7$~K at 15~GHz.  Infrared spectroscopy
indicates the source is a K5V star with faint Br $\gamma$ emission,
suggesting that it is a weak--line T Tauri object.  Zeeman splitting
measurements in the infrared spectrum find $B \sim 2.6 \pm 1.0$ 
kG.  The flare is an extreme example of magnetic activity associated with a
young stellar object.  These data suggest that
short observations obtained with ALMA will uncover hundreds of flaring
young stellar objects in the Orion region.  \end{abstract}

\keywords{open clusters and associations: individual (Orion Nebula
Cluster) --- stars: flare --- stars: formation --- stars: magnetic
fields}

\section{Introduction}

The transient radio sky is mostly unexplored, especially at millimeter
wavelengths.  The few systematic radio searches for astronomical
transients have been done at longer wavelengths, cover a small portion
of the sky, or have low sensitivity (e.g., Langston \etal 2000, Hyman
\etal 2002, Carilli \etal 2003), principally due to the lengthy observing
necessary to conduct transient surveys with even the most sensitive of
existing instruments.  Many of the known radio transients, in
fact, were first discovered at other wavelengths (e.g., Frail \etal
2003).  Therefore, the population of transient millimeter wavelength
sources is currently only poorly constrained.  With the advent of new,
more sensitive facilities (such as ALMA), we expect that discoveries of
radio transients will be a common occurrence. 

Radio stars are among the best studied transient sources at centimeter
wavelengths (G\"{u}del 2002).  A wide range of stellar types have been
shown to emit in the radio: brown dwarfs, T Tauri stars, dwarf stars,
and evolved giants.  The radio properties of young stellar objects
(YSOs) have been studied in a variety of environments, particularly in
the Orion nebula (Garay, Moran \& Reid 1987; Churchwell \etal 1987; Felli \etal 1993a, 1993b). 
These surveys show the existence of sub--arcsecond radio sources, many
of which are variable and coincident with stars.  Detection of radio
circular polarization in YSOs has strengthened the case that flares from
these stars originate in massive plasma ejections related to coronal
magnetic field activity, qualitatively similar to those seen in the Sun
(Feigelson, Carkman \& Wilking 1998, White, Pallavicini \& Kundu 1992). 
Although deep X--ray and infrared surveys of the Orion nebula have
identified many variable sources (Feigelson \etal 2002; Hillenbrand \&
Carpenter 2000; Carpenter, Hillenbrand \& Skrutskie 2001), the origins
of magnetic activity in YSOs are still poorly understood (Feigelson
\etal 2003). 

We report here the serendipitous discovery of a millimeter--wave
transient source in the Orion nebula cluster (Figure \ref{fig:bima}). 
The Berkeley--Illinois--Maryland Association Array (BIMA) detected the
source at 86~GHz on 2003 January 20, with a flux density that varied
substantially on a timescale of hours (Bower, Plambeck \& Bolatto 2003). 
For a time the object became the brightest source in the cluster, with a
flux density greater than those of the Becklin-Neugebauer Object (BN)
and the massive young star IRc2.  This detection was possible only
because BIMA was in its highest resolution configuration, which provides
a $0\farcs5$ synthesized beam and filters out the much stronger extended
emission from dust, molecular lines, and the Orion HII region. 

The source position, measured to an accuracy of $\pm 0.1$\arcsec\ from
the BIMA image, was found to be coincident with a variable
centimeter--wave radio source (GMR-A; Garay \etal 1987, Felli \etal
1993a,b), a bright near infrared point source (number 573; Hillenbrand
\& Carpenter 2000), and a variable X--ray source (number 297; Feigelson
\etal 2002).  These previous observations suggested that \object\ is a YSO
deeply obscured by the molecular cloud.  New observations presented here
support and expand this conclusion. 

Follow--up observations with BIMA and the Very Large Array (VLA) showed
that the source decayed within days of its outburst.  \object\ flared
again on several occasions over the 70 days following the first flare,
although never to the discovery intensity.  The Nobeyama Radio
Observatory (NRO) detection of \object\ at 100 and 150 GHz a few days
after discovery (Nakanishi \etal 2003) found an inverted spectrum, indicating that the
synchrotron self--absorption frequency of the plasma was greater than
100~GHz.  Circularly polarized emission was
detected with the VLA at several epochs, indicating that the radio
emission was caused by cyclotron radiation from mildly relativistic
electrons in a strong magnetic field.  Very Large Baseline Array (VLBA)
observations within 10 days of the discovery showed that a significant
fraction of the flux density of \object\ was unresolved on the
milliarcsecond scale. 

By an extraordinary coincidence, the {\em Chandra} X--ray Observatory
was in the midst of a very deep integration on the Orion nebula at the
time of the BIMA discovery (Getman \etal 2003).  The X--ray flux from
\object\ increased by a 
factor of $\sim 10$ approximately 2 days prior to the BIMA
observation.  To our knowledge, this is just the second simultaneous
radio/mm and X--ray observation of a YSO (Feigelson \etal 1994) and the
only simultaneous radio/mm and X--ray observation of a {\em flaring}
YSO. 

Infrared photometry at the Keck I 10-m and the Cerro Tololo
Interamerican Observatory (CTIO) 4-m telescopes within days of discovery
showed no variability with respect to the historical flux.  Infrared
spectroscopy with the Keck II 10-m telescope allowed a robust
identification of the infrared source as a star of spectral type K5V,
with a fast rotation, a measurable Zeeman effect indicative of magnetic
activity, and a weak Brackett $\gamma$ emission line.  All evidence
points to a strong magnetic outburst from a highly obscured young
stellar object, probably a weak--line T Tauri star. 

% This paper is organized as follows: we describe in detail the various
% observations in \S 2.  In \S 3, we discuss the identification of
% \object\ with radio, infrared and X--ray counterparts, the physical
% parameters of the flaring source and the relationship to other radio
% stars.  We also consider the number of such sources that we expect to
% find with more sensitive millimeter wavelength arrays such as CARMA and
% ALMA.  We present our conclusions in \S 4. 

\section{Observations and Data Analysis}

This paper presents observations obtained at BIMA, the VLA, the VLBA,
Keck, and the CTIO.  Plots of the light curve including dates for
infrared photometry, infrared spectroscopy and X--ray observations are
shown in Figures~\ref{fig:bimaflare}, \ref{fig:lightcurve} and
\ref{fig:lightcurvelong}.  In addition to the observations described
below, we have included the NRO observations reported by Nakanishi \etal
(2003) and the {\em Chandra} X--ray observations reported by Getman
\etal (2003).  Positions determined from various methods are listed in
Table~\ref{tab:position}.  Flux densities from BIMA, VLA, and near
infrared observations are listed in Tables~\ref{tab:bimafluxes},
\ref{tab:vlafluxes}, and \ref{tab:irphot}.  We discuss each data set in
detail below. 

\subsection{BIMA Observations}

Continuum observations of the Orion nebula at 86~GHz were made with
the BIMA array on 2002 December 23 and 2003 January 20 as part of a long
term project to track the proper motion of the Becklin Neugebauer Object
relative to IRc2 (Plambeck \etal 1995).  BIMA was in its A-configuration
(Welch \etal 1996), which provides telescope spacings of up to 1.9~km;
the synthesized beam on Orion was $0\farcs9 \times 0\farcs5$.  The
intense v=1 J=$2\rightarrow1$ SiO maser associated with IRc2 was used as
a phase reference for self--calibration. 

Figure~\ref{fig:bima} shows the discovery image, from the 2003 January
20 data set.  The map covers a $2\farcm5\times2\farcm5$ region and has
not been corrected for attenuation by the 2\farcm2 BIMA primary beam. 
The flaring object is conspicuous in the upper right quadrant.  Note
that emission from dust, molecular lines, and the Orion HII region is
almost completely resolved out in this image.  There is no sign of the
flare source in the 2002 December 23 image, to a $3\sigma$ upper limit of
6~mJy (after correction for primary beam attenuation).  BIMA observed
this field also on 2003 January 11 and 14, during very poor weather. 
Maps generated from these data show BN and IRc2 but no sign of the flare
source, to levels of 12 and 15 mJy, respectively. 

Following the preset schedule, BIMA was reconfigured from the A--array
into the B--array on 2003 January 21, so all follow--up observations of
the flare were obtained with a $5\farcs5\times2\farcs5$ synthesized
beam.  To exclude extended emission, only baselines longer than 20~$k\lambda$
were used to make these
maps.  Fortunately, the flare
source is far enough from the Orion--KL nebula that confusion from dust
or spectral line emission is unimportant at this angular resolution. 
Brief follow--up observations were conducted on 2003 January 23, 24, and
30 and 2003 February 4 and 6.  \object\ was detected only on 2003
February 6, with a flux density of $11\pm 3$ mJy. 

\subsection{VLA Observations}

VLA observations were carried out on 19 occasions beginning 2003 January
22 (two days after the discovery observations) and ending 2003 March 29. 
Before 17 February 2003 
intervals between observations ranged from one day to 5 days.
Observations had durations ranging from 1 to 3.5 hours.  They were
conducted in the standard continuum mode with two polarizations and two
IF bands of 50 MHz.  Observing frequencies of 8.4, 15, 22 and 43 GHz
were used, although not all bands were observed in each epoch. 

The data were analyzed in AIPS.  The absolute flux density scale was set
by observations of J0713+438.  Following the VLA calibrator manual, we
assume for J0713+438 $S_{8.4}=1.14$ Jy, $S_{15}=0.73$ Jy, $S_{22}=0.55$
Jy, and $S_{43}=0.29$ Jy.  These fluxes agree within 10\% of measurements
from 27 December 2002 and 8 February 2003 at 8.4, 22 and 43 GHz 
(http://www.aoc.nrao.edu/$\sim$smyers/calibration/).
Antenna pointing was stabilized using X--band
reference pointing on J0607-085.  Phase, amplitude, and leakage term
calibration were performed with observations of J0607-085.  Since these
observations are brief, the leakage term calibration cannot be regarded
as providing accuracy in the linear polarization better than 1\%. 
Instrumental errors in the fractional circular polarization are $< 1\%$ (Bower
\etal 2002). 

The VLA was in the DnC and D configuration during these observations,
providing a typical resolution of $3\arcsec \times 1\arcsec$ at 22~GHz. 
Because of the substantial extended flux associated with the Orion
nebula, only visibilities on baselines longer than $20\ k\lambda$ were
used.  The typical image rms noise in a 1--hour long observation was 10
mJy, 3 mJy, 1 mJy and 1 mJy at 8.4, 15, 22 and 43 GHz, respectively. 
Flux densities for \object\ were measured by fitting for a point source
at the phase center.  \object\ was never detected at 8.4~GHz
because confusion from extended
emission in the $10\arcsec$ synthesized beam dominated the source.  
We also measured the flux density and the position of BN at
22~GHz at each epoch.  The stability of these quantities gives us
confidence in the accuracy of our measurements for \object.  We find a
mean flux density for the BN object at 22 GHz of $12.9 \pm 0.1$ mJy. 

\subsection{VLBA Observations}

Observations with the VLBA were carried out on 2003 January 24 and 29
for 2 and 6 hours, respectively.  Observing frequencies were 15 and 22
GHz in both experiments.  The observations were performed in a
phase--referenced mode that allows detection of a weak source.  The
reference source used was J0541-0541, which is 1.6 degrees away from
\object.  We assumed a position $\alpha=$05:41:38.0833740 and
$\delta=$-05:41:49.428471 (J2000) for J0541-0541.  A cycle time (that
includes a calibrator observation, slew to the source, a source
observation and slew back to the calibrator) of 40 seconds was used in
the first epoch.  Cycle times of 60 and 120 seconds were used at 15 and
22 GHz in the second epoch, respectively.  Fringe solutions were
determined for most antennas in both experiments using the calibrator
J0541-0541.  
One iteration of self--calibration
amplitude solutions were also computed for J0541--0541, which
substantially improved the image quality.  These solutions were 
applied to \object. 

A large field was imaged and the source was clearly identified in the
second epoch at 15 and 22 GHz $\sim 40$ mas from the field center
(Figure~\ref{fig:vlba}).  Exact coordinates are given in
Table~\ref{tab:position}.  The reported errors in the position are $\sim
20 \ \mu$as in right ascension and $\sim 40\ \mu$as in
declination.  These are statistical errors.  Following Reid \etal
(1999), we estimate that systematic errors due to miscalibration of the
opacity are $\sim 100 \ \mu$as. 

During the second VLBA epoch,
the integrated flux densities measured for \object\ were $11.2 \pm 0.8$
mJy and $16.7 \pm 1.3$ mJy at 15 and 22 GHz,
respectively.  These are substantially less than the VLA fluxes measured
1 day later, of $32.4 \pm 1.7$ mJy and $30.1 \pm 0.8$ mJy at 15 and 22
GHz.  The source appears point--like in the 22 GHz image with a
resolution of $1.4 \times 0.95$ mas: its maximum deconvolved size is
$0.7$ mas.  \object\ is slightly extended in the 15 GHz image, which has
a resolution of $2.0 \pm 0.60$ mas.  At this frequency the source is
equally well--modeled as a circular Gaussian of deconvolved size 0.9
mas, or as two point--like components separated by 0.8 mas and of flux
density $8.2 \pm 0.3$ and $3.3 \pm 0.3$ mJy. 

No source was detected at either 15 or 22 GHz in the first VLBA epoch at
a level of $5\sigma$.  The rms noise in the 15 and 22 GHz images is 1.3
mJy and 1.6 mJy, respectively.  The VLA fluxes on the same day were
$18.5 \pm 1.7$ mJy and $23.7 \pm 0.6$ mJy, respectively.  A number of
factors may contribute to the nondetection of \object\ in the first
epoch, and to the discrepancy in flux density in the second epoch. 
Instrumental errors due to phase decorrelation and amplitude calibration
may contribute substantially.  Additionally, we know that
the source can vary significantly on a
time scale of hours.  Another plausible explanation is that some of the
flux is in a low-surface brightness ``halo'' which is resolved out by
the VLBA.  We discuss this possibility, which is supported by earlier
VLA observations, in more detail in \S 3.1. 

\subsection{Near Infrared Photometry}

Observations of \object\ were made on 2003 January 22 at 6 UT with the
Keck I 10-m telescope
%\footnote{The W.  M.  Keck telescope is jointly
%owned and operated by the University of California and the California
%Institute of Technology.} 
on Mauna Kea, Hawaii, and on 2003 January 24
at 1 UT with the Cerro Tololo Interamerican Observatory Blanco~4-m
telescope at Tololo, Chile. 

The Keck observations used the facility near--infrared camera (NIRC;
Matthews \& Soifer 1994), which is equipped with a $256 \times 256$
pixel Santa Barbara Research Corp.  InSb array, with a pixel size of
0.151$^{\prime\prime}$.  The images were taken at an airmass of 1.26,
and under seeing conditions of 0.9\arcsec.  We flat--fielded the data
according to standard procedures, and replaced bad pixels with the
median of surrounding pixels.  The absolute calibration of the images
was set by observing the IR standard star SJ9118 (Persson \etal 1998) at
a similar airmass (1.28).  Photometry in J, H, and K$_S$ bands was
obtained for \object\ and two other stars in the field.  We simply
integrated the intensity over the star and subtracted the average
background emission.  Corrections from K$_S$ to K band were less than
0.1 mag.

We obtained additional IR photometry from the Blanco~4-m Telescope at
the Cerro Tololo Interamerican Observatory with the facility wide-field
camera Infrared Side Port Imager (ISPI; Probst \etal 2003).  The instrument uses a Rockwell
2048$\times$2048 pixel HgCdTe detector with a 10.2\arcmin\ field of
view.  Conditions were photometric with seeing of 1.6\arcsec\
FWHM.  GMR-A was too bright at $K_S$-band to be observed in the shortest
possible integration time, so we de-focused the telescope in order to
obtain unsaturated images for this filter.  We observed the standard
star SJ~9116 from Persson et al.\ (1998) for photometric calibration,
also with the telescope defocused for $K_S$-band.  The data were reduced
in a standard fashion.  We subtracted an average bias frame from the
images.  We constructed flat fields from images of the lamp-illuminated
interior of the dome.  The individual images were registered via
cross-correlation and stacked to form a final mosaic.  Aperture
photometry was used to measure the object and standard star fluxes. 

Comparison between the Keck and CTIO results for \object\ and two
other field stars (Table~\ref{tab:irphot}) suggests that the photometric accuracy is
on the order of 0.1 mag.  

\subsection{Near Infrared Spectroscopy}

We observed \object\ with the Keck II 10-m telescope on 2003 February 6,
using the facility near--infrared echelle spectrometer (NIRSPEC; McLean
\etal 1998).  We obtained high--resolution ($R \sim 21,000$),
cross--dispersed spectra in the wavelength range 2.10--2.40 $\mu$m using
the NIRSPEC--7 order--sorting filter.  The star was imaged in two nods
along the $0\farcs432$ $\times$ $24\arcsec$ slit, separated by
$12\arcsec$ (spatial resolution was seeing limited at $\sim 1\arcsec$). 
Each integration was 300 seconds, providing a total time of 600 seconds
on the star and a signal--to--noise ratio of approximately 160.  The
spectra were dark subtracted, flat--fielded, and corrected for cosmic
rays and bad pixels.  The curved echelle orders were then rectified onto
an orthogonal slit--position/wavelength grid based on a wavelength
solution from sky (OH) emission lines and He/Ar/Ne/Kr arc lamps.  Each
pixel in the grid has a width of $\delta\lambda = 0.031$ nm.  We
subtracted sky emission by taking the difference of the two nods and
then fitting third--order polynomials to the 2D spectra
column--by--column to remove any temporal variation. 

The star's spectrum was extracted using a Gaussian weighting function
matched to the wavelength--integrated profile of each echelle order.  To
correct for atmospheric absorption, we observed a B0V star, HD~36512;
both \object\ and the calibration star were observed at an airmass of
$\sim 1.2$.  To account for photospheric absorption features and
continuum slope, the calibration star's spectrum was divided by a spline
function fit.  The resulting atmospheric absorption spectrum was then
divided into the \object\ spectra. 

It is difficult to separate stellar from nebular emission for \object. 
Br $\gamma$ emission at 2.166 $\mu$m fills the slit with a velocity
width of 29.7 km~s$^{-1}$ and an equivalent width of $\sim 6$ \AA. 
Subtraction of the mean off--source spectrum from that of \object\
reveals a weak emission line with FWHM $\sim 23$ km~s$^{-1}$ and EW
$\sim 0.8$ \AA. 

\section{Discussion}

\subsection{Radio, Infrared, and X--ray Properties}

The mm-wavelength flare source has counterparts
at radio, infrared and X--ray
wavelengths.  The object was first identified (as ``source A'') by Garay
\etal (1987) from VLA observations obtained in 1981; its flux densities
were $10.0 \pm 1.0$ mJy and $11.0 \pm 1.0$ mJy at 5 and 15 GHz,
respectively.  Later VLA observations by Felli \etal (1993a) gave flux
densities of $5.5 \pm 0.5$ mJy and $5.4 \pm 0.3$ mJy at 1.4 and 15 GHz. 
Both studies found evidence that the source was slightly extended; Felli
\etal found a deconvolved size of $0.19 \times 0.14$ arcsec.  From
observations at 13 epochs over an 8-month period in 1990, Felli \etal
(1993b) found that the flux density of \object\ was highly variable at 5
and 15 GHz, with a flat spectrum.  Our VLBA observations, in combination
with the minimum flux density in the VLA lightcurve, suggest that \object\
consists of a constant, extended, flat spectrum source with a flux
density $\sim 5$ mJy which is resolved out by the VLBA and a variable,
compact source that is detected by the VLBA. 

As shown in Figure~\ref{fig:spectra}, the radio spectral index was 
highly variable in the weeks following the initial BIMA detection.  On
22 January, 2 days after the flare was detected at BIMA, the emission
had a flat spectrum between 22 and 43 GHz ($\alpha \approx -0.1$ for
$S\propto \nu^\alpha$), but appeared to cut off sharply below 22 GHz. 
Two days later, as the first outburst cooled, the spectrum remained
peaked at roughly 22 GHz, with a higher frequency spectral index of
$-0.5$.  One day later, on 25 January, Nobeyama observations (Nakanishi
\etal 2003) detected the source at 98 and 147~GHz; the spectrum was
inverted ($\alpha = 1.4$), implying that the synchrotron self-absorption
frequency of the plasma was greater than $\sim 100$~GHz.  On 7 February
the source appeared to have a flat spectrum from 15 to 86 GHz.  Much
later, on 27 March, the spectral index is $-0.7$, although the absence
of an 86 GHz observation makes this steep spectrum somewhat
uncertain. 

Circular polarization is clearly detected in many epochs
(Table~\ref{tab:vlafluxes}).  When either the circular polarization or the
total intensity is not detected at the $2-\sigma$ level, we report
an upper limit to the absolute value of the circular polarization.  The mean
fractional circular polarization (measured as $V/I$) 
is $-3.4 \pm 0.4\%$, $-3.9 \pm 0.2\%$ and $-6.6
\pm 0.6\%$ at 15, 22 and 43 GHz, respectively.  Nevertheless, there is clear
time variability in the circular polarization: in particular, we did not
detect circular polarization near the peak of the flare.  {\em The
emission is never linearly polarized.} We searched for linear
polarization but did not detect it in any epoch with limits $\sim 1\%$. 

The near infrared counterpart of \object\ is bright, and not variable. 
\object\ is positionally coincident with source \#573 in the infrared
survey of Hillenbrand and Carpenter (2000).  Our infrared photometry
gives H and K magnitudes consistent with those measured by Hillenbrand and
Carpenter (Table~\ref{tab:irphot}).  Furthermore, the intensive survey
for variability in the Orion nebula by Carpenter, Hillenbrand \&
Skrutskie (2001) did not detect this object as significantly variable. 

In contrast, the X--ray counterpart of \object\ is characterized as a flare
source by Feigelson \etal (2002);
it is substantially variable on a timescale of
less than 12 hours as well as over six months.  The spectrum is
consistent with an intrinsic X--ray luminosity $L_x=10^{31.7}$ erg
s$^{-1}$ attenuated by a gas column density $N_H=10^{22.6}$ cm$^{-2}$. 
This luminosity ranks it among the brightest 10\% of the X-ray sources
in the Orion nebula.  Getman \etal (2003) reported a significant X--ray
flare ($\times 10$) at its position.  Their preliminary light curve shows 
the flare beginning $\sim 2$ days
before the detected millimeter wave flare, and continued through the time of the
radio flare.  Of course, we are not able to accurately determine the onset of the flare 
at millimeter wavelengths because of the scarcity of observations.  The X--ray
flux was declining at the time of the millimeter flare.
Modeling of the X--ray spectrum by Getman \etal during the flare indicates
an absorption column density comparable to that found earlier. 

\subsection{Variability Timescale of the Radio Emission}

To study the flux variability during the onset of the radio flare, we
broke up the 8--hour long visibility data obtained by BIMA on the night
of 2003 January 20 into 1--hour blocks.  The flare brightened by a
factor of $\sim 4$ to 8 during the track (Figure~\ref{fig:bimaflare}),
while the fluxes of IRc2 and BN remained stable.  The flux density of
the flare source was fit in each interval after subtracting the
contributions from point sources at the positions of IRc2 and BN with
flux densities of 45 and 68 mJy respectively.  Because the BIMA
receivers are sensitive to linear polarization in the vertical
direction, a concern is that the observed variability may not be
intrinsic, but introduced by linearly polarized emission combined with
changes in the parallactic angle during the night. The parallactic angle
changed from $-38^\circ$ to $+45^\circ$ during the observations.  If
this were the cause of the variability, the emission would have to be
$>45\%$ linearly polarized, which is highly unusual.  A broad range of
position angles and polarization fractions were used to model the light
curve, obtaining a minimum reduced $\chi^2\sim17$, suggesting that
linearly polarized emission cannot reproduce the observations. 
Furthermore, linear polarization is undetected in all subsequent VLA
observations.  It is very unlikely that the 86 GHz flux is initially
strongly polarized while two days later the 43 GHz flux is not.  This
fact, together with the poor quality of the fit, strongly argue that the
observed 86~GHz lightcurve is due to intrinsic variability and not to
polarization. 

We searched for short--term variations in the VLA data, as well.  Most
of the observations were split into two or more blocks probing
timescales that range from 0.5 to 1.5 hour.  There was no evidence for
significant variability on these timescales. 

\subsection{Identification of \object\ as a Young Stellar Object}

To determine the spectral type of \object, we compared its near--IR
spectrum with the medium--resolution $K$--band spectral atlas of Wallace
\& Hinkle (1997).  The relative strengths of the Al and Mg lines near
2.11 $\mu$m, as well as the relative strengths of the Ti and Si lines
near 2.18 $\mu$m, indicate a spectral type of K5.  The shallow CO
bandheads are consistent with a dwarf star.  On this basis, we identify
\object\ as a spectral type K5V. 

The $^{12}$CO (v=$0\rightarrow2$) rovibrational band (extending redward
from the bandhead at 2.29 $\mu$m) allows us to measure the radial
and rotational velocities of the star, $v_{lsr}$ and $v \sin i$,
respectively.  As a template for our rotational model, we used a
high--resolution FTS sunspot umbral spectrum (Wallace \& Hinkle 2001)
with $T_{eff} = 4100$K.  We convolved the sunspot spectrum with a
rotational broadening profile from Gray (1976), and convolved the result
with the measured profile of an OH line observed with
NIRSPEC to account for the instrumental resolution.  Iterating over a range of
rotational velocities, we find $v \sin i = 23 \pm 2$ km~s$^{-1}$ and
$v_{lsr} = -4 \pm 5$ km~s$^{-1}$.  The quoted error in $v\sin i$
reflects the broad nature of the minimum in the residuals for models
near 23 km s$^{-1}$.  Our determination of 
$v_{lsr}$ suggests that \object\ is not embedded in the
molecular cloud, for which the radial velocity from NH$_3$ measurements is
$v_{lsr}=9 \pm 1$ km s$^{-1}$ (Batrla \etal 1983), but rather lies behind it. 

Despite the difficulties in subtracting the background emission from the
nebula, the observed Br $\gamma$ emission is clearly substantially
weaker and narrower than is typical in classical T Tauri stars (Folha \&
Emerson 2001), indicating that this is probably a weak line T Tauri star
(WTTS).  This designation is also consistent with strong radio
variability, as WTTS are the most luminous and most variable of radio YSOs.  

Taken together, the X--ray data, IR photometry, and IR spectroscopy
point to a clear identification of \object\ as a YSO inside, or perhaps
behind, a molecular cloud.  
However, the identification of \object\ as a K5V is not consistent with the high
luminosity implied by the very bright K magnitude and 
the large extinction.  We discuss here the different scenarios that might 
resolve this, including different degrees of extinction, the presence of an 
infrared excess, and classification as a K5III star, but none of these are fully 
satisfactory.  In the simplest
approach, we use the X--ray absorption column depth and $R_V=3.1$ to
find the extinction $A_V=21.4$ which implies $A_K = 2.4$ mag (Rieke \&
Lebofsky 1985; Binney \& Merrifield 1998).  Using color and bolometric
corrections appropriate for K5V stars, we find $L \sim 30 L_\sun$.
But this result is almost certainly wrong
for two reasons:  (1), this luminosity exceeds the luminosity of the stellar
birth line, which is $<10 L_\sun$ at this
effective temperature (Palla \& Stahler 1999); and (2), the photospheric H--K color
of 1.0 after applying reddening correction is not appropriate for a K5V star.
Assuming a photospheric color H--K=0.1,
which is the case for K5V stars, we find $A_V=35$ and
$A_K=3.9$.  A similar extinction is obtained with examination of the J--H and H--K
colors (e.g., Haisch, Lada \& Lada 2000).  
While this satisfies the spectroscopic constraints, it
produces a luminosity of 100 $L_\sun$, again in excess of the stellar birth line.
This inconsistency suggests an excess of infrared
emission possibly due to a disk.  Assuming the reddening calculated by
the X--ray absorption column depth ($A_K=2.4$) and a photospheric color H--K=0.1,
then we find an H--K excess of 0.9 mag and a K--band excess of 1.7 mag
(Hillenbrand \& Carpenter 2000).  This reduces the bolometric luminosity
to 6 $L_\sun$.  While this satisfies
the stellar birth line luminosity limit and the colors for these stellar
classes, the presence of a strong infrared excess 
is more consistent with a classical T Tauri star than with a 
WTTS, although some WTTS do have an IR excess (Strom \etal 1989).
Considering the same scenarios for K5 giant or supergiant stars leads
to the same problems.  

Could \object\ be something other than a T Tauri star?  A possibility is
that the flaring source active in radio and X--rays is not the object
dominating the IR luminosity, hence the lack of infrared activity. 
There is some precedent for faint companion stars producing intense
variable radio emission: astrometric radio and optical measurements of
$\theta^1$~Ori~A have shown that the object responsible for radio
flaring is actually part of a triple system (Garrington \etal 2002). 
The ``active'' companion, however, would have to be several magnitudes
fainter than the main K5V star even during its flaring state, to account
for its non detection in the IR spectrum.  Adaptive optics observations
in the infrared would be necessary to identify any faint companions (e.g.,
Duch\^{e}ne \etal 2003).  Other
potential radio transient source types are brown dwarfs and late--type
main sequence stars (e.g., Berger 2002).  These sources, however, are too
faint by many orders of magnitude to account for the quiescent or
flaring radio emission from \object.  Evolved stars such as RS Cvn binaries
are unlikely to be 
associated with the Orion complex.  In the following
section we show that the magnetic field measurements also support the
conclusion that \object\ is a T Tauri object. 

\subsection{Zeeman Measurement of Magnetic Field Strength}

We used the echelle infrared spectrum to search for Zeeman broadening of
lines in the stellar photosphere, which would indicate the presence of a
strong magnetic field.  We selected the strong Ti{\small I} lines at
$\lambda=2.17886\ \mu$m and $\lambda=2.19033\ \mu$m present in echelle
order 35 in our spectra to carry out this study.  These transitions have
spectroscopic terms (3d$^3\,^5$P$_3\rightarrow$4p$^1\,^5$D$_4$) and
(3d$^3\,^5$P$_2\rightarrow$4p$^1\,^5$D$_3$) respectively, resulting in
very similar effective Land\'e $g$--factors, 1.667 and 1.833.  Because
the $g_L$ factors of the upper and lower energy levels in either
transition are not precisely equal, these lines formally exhibit the
``anomalous'' Zeeman effect and they are broken up into seven components
by the magnetic field.  In practice, since the Land\'e factors are
almost equal, the emission is separated into approximately 3 equal
components. 

We modeled the observed emission using 3 spectral lines with the shape of the
intrinsic instrumental profile convolved with the rotational profile
corresponding to the measured value of $v\sin i=23$ km s$^{-1}$.  Thus,
we assume that the only sources of line broadening are instrumental,
rotational, and Zeeman, an assumption that seems justified in view of
the large $v\sin i$.  These models were computed for a grid of Zeeman
splittings, $\Delta\lambda$, and line amplitudes.  We determined the
best model by evaluating the reduced $\chi^2$ of the residuals.  The
error was estimated by finding the values of the parameters that
increased the reduced $\chi^2$ by one.  Figure \ref{fig:zeeman} shows
the results of the fit for the $\lambda=2.17886 \mu$m transition.  The
measured Zeeman splittings are $\Delta\lambda=0.14\pm0.07$ nm ($\chi^2$
= 1.4) and $\Delta\lambda=0.09\pm0.05$ nm ($\chi^2$ = 1.7),
corresponding to magnetic fields $3.8\pm2$ kG and $2.2\pm1.1$ kG.  Thus,
our best estimate of the photospheric magnetic field traced by the
Ti{\small I} lines is $2.6\pm1.0$ kG.  Observations of T Tauri objects
find similar magnetic field strengths (e.g., Johns--Krull, Valenti \&
Koresko 1999). 

\subsection{Radio Emission Mechanism}

Several lines of evidence indicate that the radio emission mechanism is
nonthermal: the compactness of the emission detected using the VLBA, the
presence of circular polarization, the strongly variable spectral index,
and the rapid rise and decay time of the total intensity.  We can set a
lower limit to the brightness temperature based on the flux measured by
the VLBA observations.  We find $T_b \ga 5 \times 10^7$ K and $T_b \ga 3
\times 10^7$ K at 15 and 22 GHz.  Electrons in equilibrium with this
temperature will have velocity $\ga 0.07c$, or an electron Lorentz factor
$\gamma_e \ga 1.003$. 

The presence of high fractional circular polarization rules out thermal
bremsstrahlung and synchrotron radiation as emission mechanisms, and
indicates that the dominant emission is cyclotron radiation generated by
mildly relativistic electrons in the presence of strong magnetic fields
(e.g., G\"{u}del 2002).  The observed mean fractional circular polarization in the
VLA bands is $\sim -3$ to $-7\%$, with peak values greater than $-10\%$. 
The absence of linear polarization also argues against synchrotron
emission, although very tangled magnetic fields or a high rotation
measure due to an intervening dense, magnetized plasma could
depolarize a synchrotron source. 

The radio emission probably originates in coronal magnetic fields,
which are an order of magnitude weaker than the photospheric magnetic
field strength determined in the previous section.  Assuming a field
strength for the radio emission of 100 G, the corresponding
electron gyro--frequency is $0.5\gamma_e^{-1}$ GHz. 
The lack of a very steep fall--off at high frequencies, characteristic
of a single temperature thermal cyclotron emitter, leads us to conclude
that the emission originates from a power--law distribution of electrons
with low energy.  A power--law distribution of energies with index
$\delta=1.5$ yields an optically--thin spectrum with $\alpha=-0.1$ above
a peak frequency of 22 GHz.  The characteristic size of the emitting region for the
assumed magnetic field is $2\times 10^{11}$ cm.  For comparison, models indicate
that the radius of a YSO with $L=3.5 L_\sun$ and $T_{eff}=4100$ K is
$\sim 4\times10^{11}$ cm, suggesting that the cyclotron emission is a
global phenomenon for the star. 

A significant anomaly in this picture is the absence of circular
polarization in the first VLA observation on January 22, 2 days after the
millimeter wavelength flare.  A possible
mechanism to initially supress circular polarization is to invoke an
initial outburst of relativistic electrons, leading to synchrotron
rather than cyclotron emission.  Rapid cooling through expansion or
X--ray radiation would then drive the radio source into the cyclotron
regime.  In this scheme one expects that the early radiation would be
linearly polarized.  We have seen that VLA observations on day
22 failed to show any linear polarization.  As discussed above, however,
there are mechanisms that could significantly depolarize a synchrotron
source.  An alternative explanation for the initial lack of circular
polarization assumes the existence of two oppositely polarized sources
associated with the star.  Imaging of some flaring radio stars has shown
such oppositely polarized lobes (Mutel \etal 1998).  Circular
polarization disappears when these lobes are equally balanced and
appears when one lobe comes to dominate the emission.  Without more
detailed modeling and/or imaging of the source we cannot discriminate
between these two possibilities.

The source reflared several times during the course of the follow--up
observations, albeit never to the level of the discovery data.  In fact,
the light curve appears to consist of multiple flares spread over $~\sim
70$ days, with individual flares having a rise and decay timescale of
days or shorter.  During the BIMA discovery, we measure a rise time of
$\sim 1$ hour.  Since this is the duration of most of the VLA
observations, we cannot clearly discern whether there is a similar
variability timescale in those data.  The timescales for radiation
losses for a cyclotron source are proportional to $B^{-2}$, and are
$\sim 1$ day for $B=100$ G, which is comparable to what is observed.

\subsection{The Radio to X--ray Luminosity Correlation}

The luminosity of this flare is comparable to that of the brightest
stellar radio flare ever detected.  The peak luminosity at 86 GHz
during the flare was $4\times 10^{19} {\rm\ erg\ s^{-1} Hz^{-1}}$,
only 50\% less than the peak flare of the current record--holder, the
FK Com--type giant star HD32918 (Slee \etal 1987, Bunton \etal 1989).
The luminosity of the \object\ flare, however, was more than an order
of magnitude higher than detected for any YSO.  Typical maximum radio
luminosities for YSOs are $\sim 10^{18} {\rm\ erg\ s^{-1} Hz^{-1}}$
(G\"{u}del 2002).  Since \object\ flared at millimeter wavelengths,
its total luminosity was $L_\nu \sim 3 \times 10^{30} {\rm\ erg\
s^{-1}}$, about two orders of magnitude higher than that of centimeter
wavelength sources.

The ratios of X--ray and radio emission for \object\ during both its
quiescent and flaring states approximately follow the correlation
found for active stars and solar flares (G\"{u}del 2002)
\begin{equation}
{L_x \over L_r} \approx 10^{15 \pm 1} [Hz].
\end{equation}
This correlation was derived from radio observations at centimeter
wavelengths (typically 8 GHz): the precise value of the coefficient
is unknown for measurements at millimeter wavelengths.  To compute the
luminosities, we adopted the quiescent, absorption--corrected X--ray
luminosity $L_x=10^{31.7}$ erg\ s$^{-1}$ measured by {\em Chandra}.
During the flaring state the {\em Chandra} observations show the count
rate increasing by a factor $\sim 10$, leading to a prediction of $L_r
\approx5\times10^{18\pm1} {\rm\ erg\ s^{-1} Hz^{-1}}$, very similar
to that observed at 86 GHz.  These data support the conclusion that
the flare is caused by magnetic activity common to other radio stars.

\subsection{Rate of Flaring Activity in the Orion Nebula}

We can parametrize the rate of millimeter wavelength flares in the Orion
nebula as 
\begin{equation}
N(S > S_0) = A ({S_0 \over 100{\rm\ mJy}})^{-\alpha}.
\label{eqn:ns}
\end{equation}
Nonthermal activity often follows a power--law index $\alpha=1$.  This
is the first flare brighter than 100 mJy detected by BIMA in $\sim 40$
observations of the Orion nebula, setting $A=0.01 - 0.1 {\rm\
day^{-1}}$.  This value of $A$ is consistent with the 22 and 43 GHz
observations described in this paper, as well as the 15 GHz dataset of
Felli \etal (1993b).  Both of these datasets show that \object\ has a
flux density of 10 mJy about 10 times as frequently as it has a flux
density $\sim 100$ mJy.  Following Equation \ref{eqn:ns}, observations
of Orion at a 0.1 mJy detection threshhold will find $\sim 10 - 100$
flares from YSOs per observation.  In fact, the monitoring campaign of
Felli \etal (1993b), with a sensitivity ranging between 0.2 and 1.2 mJy,
identified 13 variable sources.  The proposed Atacama Large Millimeter
Array will achieve a sensitivity of $0.1$ mJy or better in a 1 minute
integration (http://www.alma.nrao.edu).  Thus, a short ALMA observation
with a sensitivity of 10 $\mu$Jy may find $\sim 100 - 1000$ flares from
YSOs, which is an appreciable fraction of the total number of objects in
the region. 

The same argument can be made using the X--ray luminosity
function.  With a radio sensitivity of $3 \times 10^{15} {\rm\ erg\
s^{-1}\ Hz^{-1}}$ (equivalent to $10 \mu{\rm Jy}$), the $L_x/L_r$
correlation predicts that we will detect all stars with $L_x>
10^{30.5\pm 1}$.  This is consistent with the mean
$\log{L_x}=29.4 \pm 0.7$ of the {\em Chandra} sample, suggesting that
we would detect half of all X--ray objects in the Orion nebula
(Feigelson \etal 2002).  Approximately half of the 1075 objects in the
{\em Chandra} sample are identified as variable on short and/or long
timescales, indicating again that ALMA can expect to find hundreds
of variable objects in a few hours.

Variable sources violate the assumption of a constant sky that
underlies synthesis imaging.  The potentially large number of variable
sources in the Orion nebula and other star-forming regions introduces
a complication not yet considered for imaging with ALMA.  If not
adequately accounted for, these variable sources will introduce a
dynamic range limit to images.  Consider a source with a flare of flux
density $S_{max}$ over the quiescent flux density of $S_q$ that lasts
for $t_{max} << t_{obs}$, where $t_{obs}$ is the total length of the
observation.  An array will have a peak fractional sidelobe level
$\eta_{max}$ that is a function of array configuration, source
position and $t_{max}$.  To first order, this will introduce an
imaging error $\eta_{max} S_{max}$ after deconvolution (e.g., using
CLEAN).  For ALMA, $\eta \sim 2\%$ for observations of length $>$ 1
hour (e.g., Boone 2002).  Thus, a brief 1 mJy flare will introduce an
error $>20$ $\mu$Jy in an image that could have a theoretical rms of
10 $\mu$Jy.  Multiple sources and multiple flares complicate this
picture but are unlikely to reduce the error.  The solution, in
principle, is to image and deconvolve on a timescale comparable to the
flaring timescale.

\section{Conclusions}

We have described here the serendipitous discovery and subsequent
follow--up observations of a flaring millimeter wavelength source in
the Orion star--forming cluster.  The source appears to be a weak-line
T Tauri star highly obscured by dusty material.  Although previously
known as a variable cm radio source (GMR-A), the magnitude of the flare
and the measurement of its spectrum to millimeter wavelengths are
novel.  This is one of the most luminous radio flares observed, and
the most luminous flare from a YSO.  Future observations with greater
sensitivity can be expected to detect many more of these objects.

The discovery of a compact source, detectable by the VLBA, in
the Orion nebula raises the possibility of the measurement of
trigonometric parallax to this star--forming region.  It would be the
most accurate distance measurement for Orion, achieving a parallactic
error $\sim 100 {\rm\ \mu as}$ equivalent to a distance error of
$\sim 2\%$ with only a few measurements.  If the quiescent source is
detectable, or if the source flares again, then two more observations
are sufficient to separate the proper motion and parallax of this
source.  Monitoring and detection of other transient sources in this
region and other galactic star--forming regions could significantly
reduce the uncertainty in key star--forming parameters.  A statistical
approach is possible through VLBI astrometry of the $>10$ other
nonthermal sources in Orion previously identified by Felli \etal
(1993a).  Finally, this may allow a 3--dimensional probe of the
positions and velocities in the star forming cluster, giving vital
clues to its history.

\acknowledgements 

The BIMA array is operated by the
Berkeley--Illinois--Maryland Association under funding from the National
Science Foundation.  The National Radio Astronomy Observatory is a
facility of the National Science Foundation operated under cooperative
agreement by Associated Universities, Inc.  We thank Barry Clark for
quickly and frequently scheduling follow--up VLA and VLBA observations
of this source.  
The Keck data presented in this paper were obtained at the W. M. Keck
Observatory, which is operated as a scientific partnership among the
California Institute of Technology, the University of California and the
National Aeronautics and Space Administration. The Observatory was made
possible by the generous financial support of the W. M. Keck
Foundation. The authors wish to recognize and acknowledge the very significant
cultural role and reverence that the summit of Mauna Kea has always
had within the indigenous Hawaiian community.  We are most fortunate
to have the opportunity to conduct observations from this mountain.
It is a pleasure to thank Ron Probst,
Nicole van der Bliek, and Angel Guerra for assistance with the CTIO
observations.  M. Liu is grateful for research support from the Beatrice
Watson Parrent Fellowship at the University of Hawaii, the AAS Small
Research Grants program, and NASA grant HST-HF-01152.01-A.
We also thank Eric Feigelson, Konstantin Getman and
their collaborators for promptly sharing {\em Chandra} data on this
object, as well as their insights.

%\bibliographystyle{apj}
%\bibliography{../../myrefs}

\section{References}

\noindent
%Barrado y Navscu\'{e}s, D. \etal, 2003, \aap, 404, 171 \\
Batrla, W., Wilson, T.L., Bastien, P. \& Ruf, K., 1983, \aap, 128, 279 \\
Berger, E., 2002, \apj, 572, 503 \\
Binney, J. \& Merrifield, M., 1998, Galactic Astronomy, Princeton University Press, Princeton \\
Boone, F., 2002, ALMA Memos \#400 \\
Bower, G.C., Falcke, H., Sault, R.J. \& Backer, D.C., 2002, ApJ, 571, 843 \\
Bower, G.C., Plambeck, R.L. \& Bolatto, A., 2003, IAUC, 8055 \\
Bunton, J. D., Large, M. I., Slee, O. B., Stewart, R. T., Robinson, R. D., \& Thatcher, J. D., 1989, PASA, 8, 127 \\
Carilli, C., Ivison, R. \& Frail, D.A., 2003, \aj, in press \\
Carpenter, J.M., Hillenbrand, L.A. \& Skrutskie, M.F., 2001, \aj, 121, 3160 \\
Churchwell, E., Felli, M., Wood, D.O.S., \& Massi, M. 1987, \apj, 321, 516 \\
%D'Antona, F. \& Mazzitelli, I., 1994, \apjs, 90, 467 \\
%Dhawan, V., Mirabel, I. F. \& Rodríguez, L. F., 2000, \apj, 543, 373 \\
Duch\^{e}ne, G., Ghez, A.M., McCabe, C. \& Weinberger, A.J., 2003, \apj, 592, 288 \\
Feigelson, E.D. \etal, 1994, \apj, 432, 373 \\
Feigelson, E.D., Carkner, L. \& Wilking, B.A., 1998, \apjl, 494, L215 \\
Feigelson, E.D. \etal, 2002, \apj, 574, 258 \\
Feigelson, E.D. \etal, 2003, \apj, 584, 911 \\
Felli, M., Churchwell, E., Wilson, T.L. \& Taylor, G.B., 1993a, \aaps, 98, 137 \\
Felli, M., Taylor, G.B., Catarzi, M., Churchwell, E. \& Kurtz, S., 1993b, \aaps, 101, 127 \\
Folha, D.F.M. \& Emerson, J.P., 2001, \aap, 365, 90 \\
%Frail, D. \& Kulkarni, 1997, IAUC 6662 \\
Frail, D.A., Kulkarni, S.R., Berger, E. \& Wieringa, M.H., 2003, \aj, 125, 2299\\
Garay, G., Moran, J.M. \& Reid, M.J., 1987, \apj, 314, 535 \\
Garrington, S. \etal, 2002, 6th Euro. VLBI Net. Symp., Ros, E., Porcas, R.W., Lobanov, A.P \& Zensus, J.A., eds. \\
Getman, K.V., Feigelson, E.D., Garmire, G., Murray, S.S., Harnden, F.R.,Jr. , 2003, IAUC, 8068 \\
Gray, D.F., 1976, Wiley-Interscience, New York \\
G\"{u}del, M., 2002, \araa, 40, 217 \\
Haisch, K.E., Lada, E.A. \& Lada, C.J., 2000, \aj, 120, 1396 \\
Hillenbrand, L.A. \& Carpenter, J.M., 2000, \apj, 540, 236 \\
Hyman, S. D., Lazio, T. J. W., Kassim, N. E. \& Bartleson, A. L., 2002, \aj, 123, 1497 \\
Johns-Krull, C.M., Valenti, J.A. \& Koresko, C., 1999, \apj, 516, 900 \\
%Kasper, M.E. \etal, 2002, \apj, 568, 267 \\
Langston, G., Minter, A., D'Addario, L., Eberhardt, K., Koski, K. \& Zuber, J., 2000, \aj, 119, 2801 \\
%Levinson, A., Ofek, E. O., Waxman, E. \& Gal-Yam, A., 2002, \apj, 576, 923 \\
Matthews, K., and B.T. Soifer, 1994, in {\it Infrared Arrays in
Astronomy: The Next Generation} (I. S. McClean, Ed.), pp. 239-246,
Kluwer, Dordrecht \\
McLean, I.S. \etal, 1998, SPIE, 3354, 566 \\
Mutel, R.L., Molnra, L.A., Waltman, E.B. \& Ghigo, F.D., 1998, \aj, 93, 1220 \\
Nakanishi, K., Saito, M. Furuya, R.S., Shinnaga, H. \& Momose, M., 2003, IAUC, 8060 \\
Palla, F. \& Stahler, S.W., 1999, \apj, 525, 772 \\
Persson, S. E., D.C. Murphy, W. Krzeminski, M. Roth, and
Plambeck, R. L., Wright, M. C. H., Mundy, L. G., \& Looney, L. W., 1995, \apjl, 455, L189 \\
Probst, R.G. \etal, 2003, SPIE, 4841, 411\\
M.J. Rieke, 1998, \aj, 116, 2475 \\
Reid, M. J., Readhead, A. C. S., Vermeulen, R. C., \& Treuhaft, R. N., 1999, \apj, 524, 816 \\
Rieke, G.H., \& Lebofsky, M.J. 1985, ApJ, 288, 618 \\
Slee, O.B. \etal, 1987, \mnras, 227, 467 \\
Strom, K. M., Strom, S. E., Edwards, S., Cabrit, S., \& Skrutskie, M. F. 1989, \aj, 97, 1451 \\
Wallace, L. \& Hinkle, K., 1997, \apjs, 111, 445 \\
Wallace, L., Hinkle, K. \& Livingston, W.C., 2001, NSO Tech. Rep.  \#01-001; Tucson: National Solar
Observatory \\
%Weiler, K. W., Panagia, N., Montes, M. J. \& Sramek, R. A., 2002, \araa, 40, 387 \\
Welch, W. J. \etal, 1996, \pasp, 108, 93 \\  
White, S.M., Pallavicini, R. \& Kundu, M.R., 1992, \aap, 259, 149 \\

\newpage
\plotone{f1.eps} \figcaption[f1.eps]{BIMA 86~GHz
continuum image of Orion--KL taken on 2003~January~20, showing the
flare source at upper right.  The synthesized beam is $0\farcs93\times
0\farcs56$ at P.A. 43$^\circ$.  The halftone scale ranges from 12 to 60
mJy beam$^{-1}$.  The map has not been corrected for attenuation by the
$2\farcm2$ FWHM primary beam, shown by the circle.  Source I, the
radio source associated with IRc2, is at the center the map.  The
faint arc just below source I is dust emission from the Orion hot
core.  The Becklin--Neugebauer object is $10\arcsec$ NW of source I.
\label{fig:bima}}

\plotone{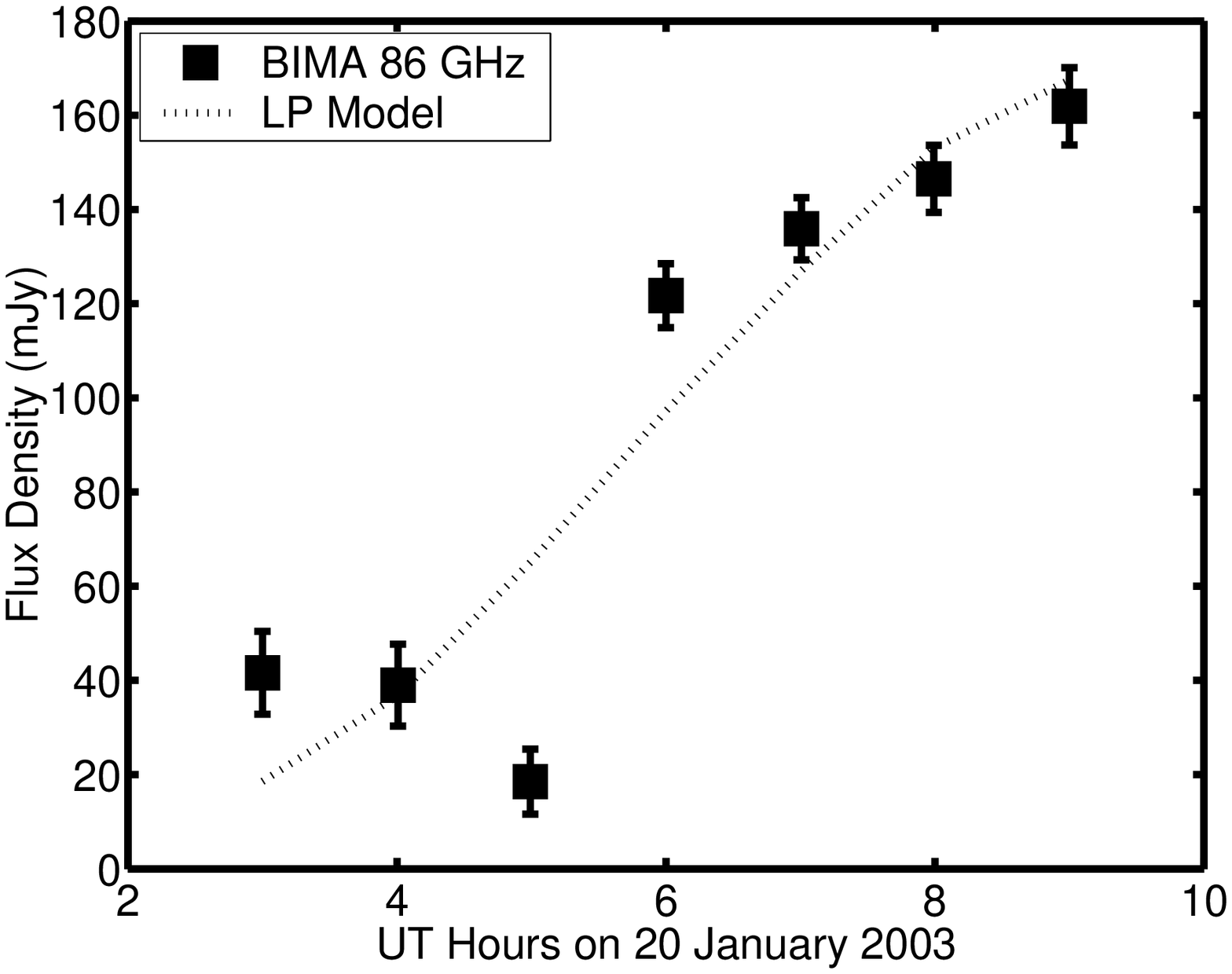} \figcaption[f2.eps]{BIMA flux densities
at 86 GHz with a model of linearly polarized flux assuming $p=50\%$
and a position angle of 107 degrees.  While the polarization model
follows the trend of the data, it is not an adequate fit for any
position angle and polarization fraction.
\label{fig:bimaflare}}

\plotone{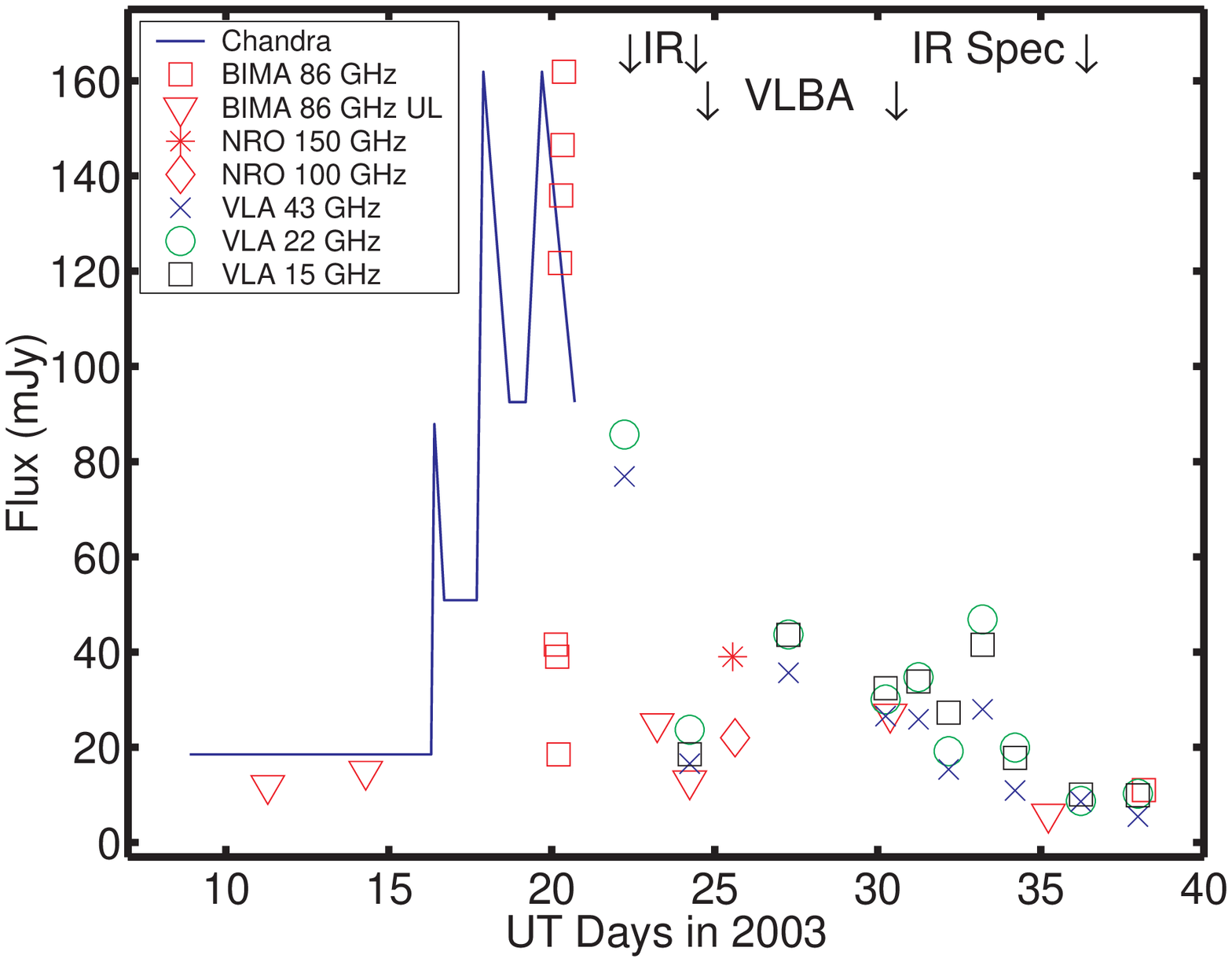}
\figcaption[f3.eps]{Light curve at millimeter,
radio and X--ray wavelengths showing early evolution of the flare along
with dates of infrared photometry (``IR'' arrows), VLBA observations
(``VLBA'' arrows) and infrared spectroscopy (``IR Spec'' arrow).
Symbols are as given in the legend.  
The downward-facing red triangles (``BIMA 86 GHz UL'')
indicate $3-\sigma$ upper limits.  
The X--ray light curve is from the
description in Getman \etal (2003) and is given in arbitrary flux units.
\label{fig:lightcurve}}

\plotone{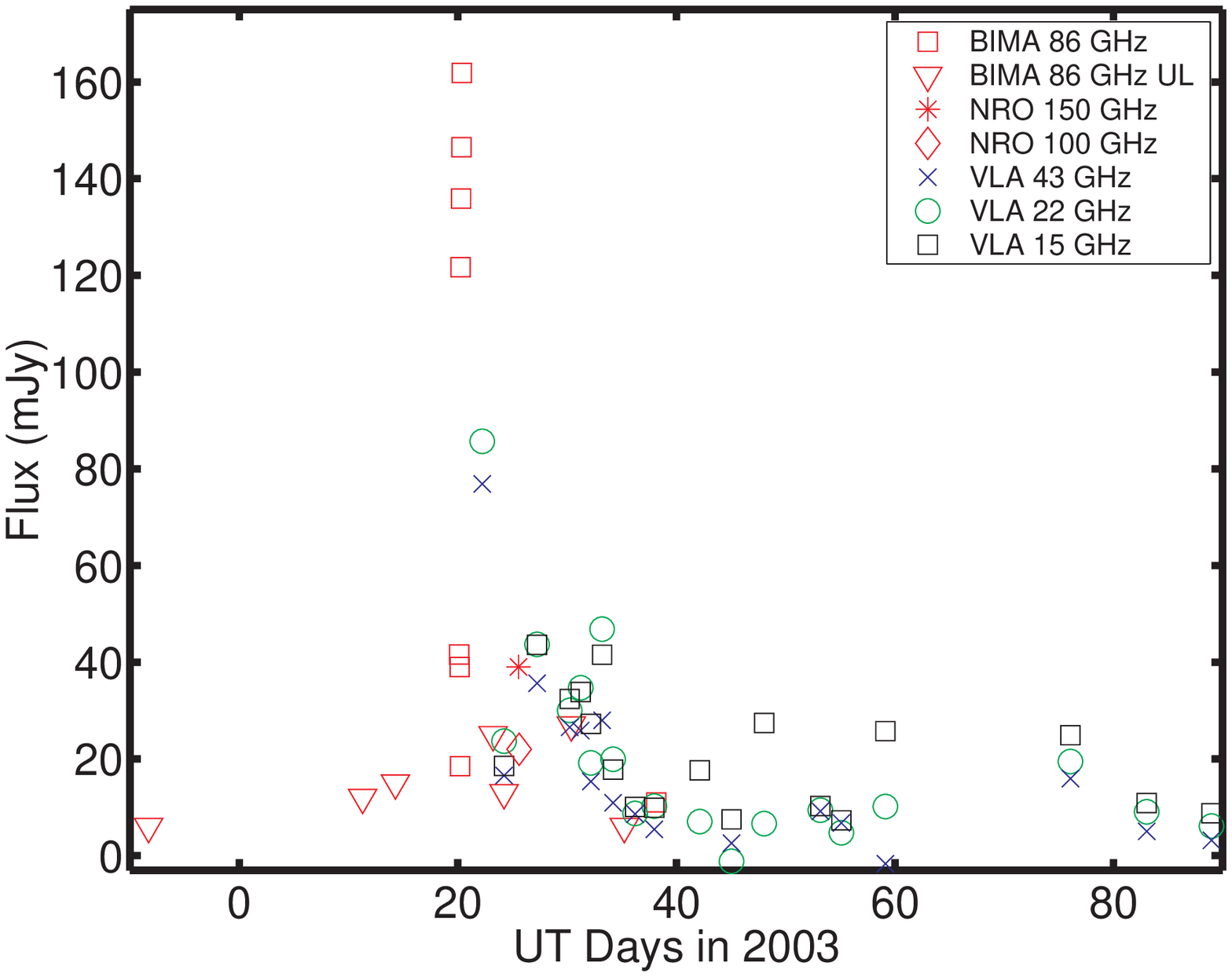} \figcaption[f4.eps]{Full
light curve at millimeter and radio wavelengths showing the long term
evolution.  Symbols are as given in the legend.
The downward-facing red triangles (``BIMA 86 GHz UL'')
indicate $3-\sigma$ upper limits.  
\label{fig:lightcurvelong}}

\plotone{f5.eps} \figcaption[f5.eps]{VLBA images of
\object\ at 22 GHz obtained on 2003 January 29.  Contours are -3, 3,
6, 12, and 24 times the rms noise levels of 0.70 mJy beam$^{-1}$ at 22
GHz.  The synthesized beam is shown in the lower left hand corner of
the image.
\label{fig:vlba}}

\plotone{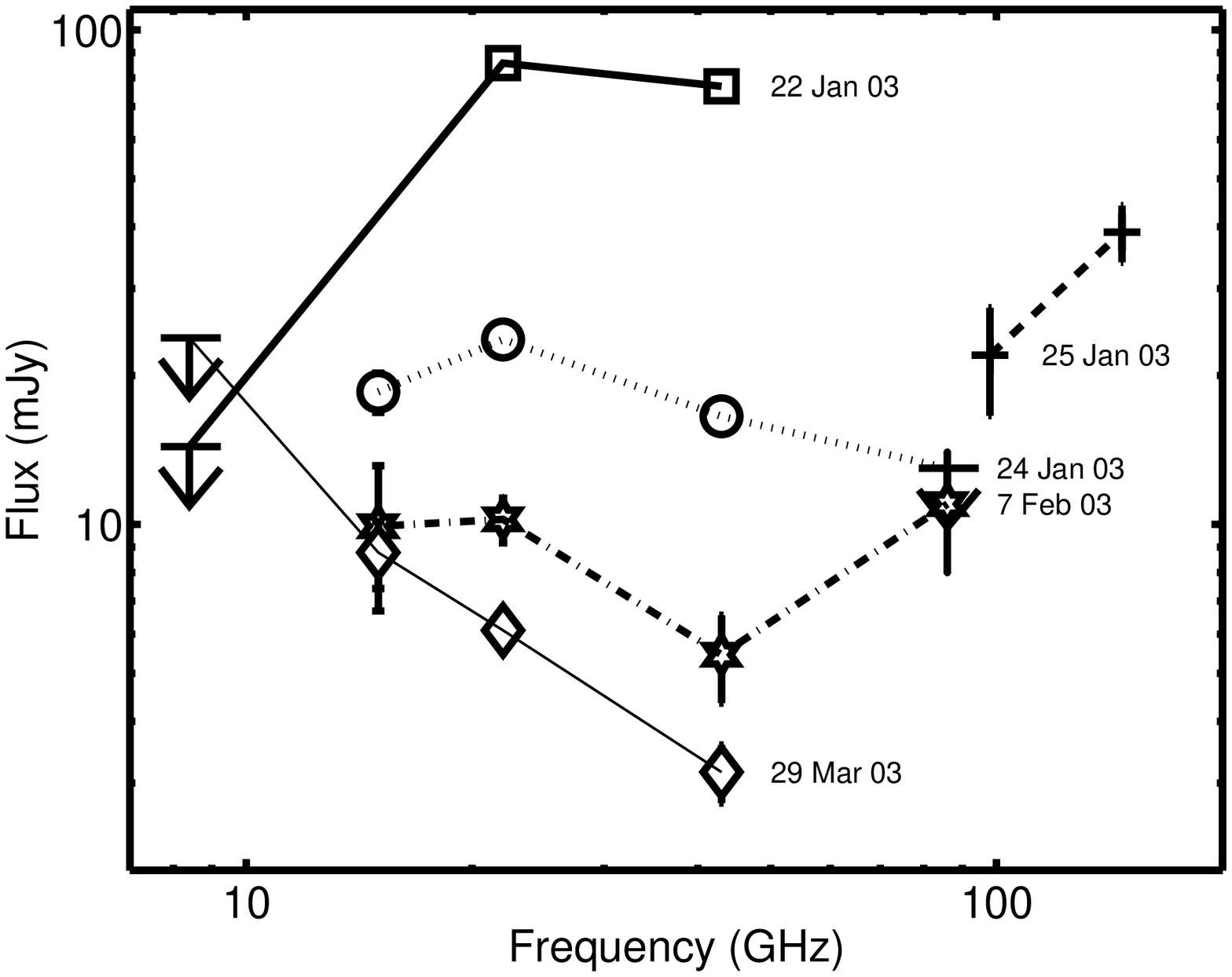} \figcaption[f6.eps]{Spectra of \object\
obtained on 2003 January 22 (squares, heavy solid line), January 24 (circles, dotted line), 
January 25 (crosses, dashed line), 
February 17 (stars, dot-dashed line) and March 29
(diamonds, light solid line).  Upper limits are $3-\sigma$.  
$1-\sigma$ errors are plotted where they
exceed the symbol size.  The 97 and 148 GHz data 
of 25 January 2003
are from Nakanishi \etal (2003).
\label{fig:spectra}}

\plotone{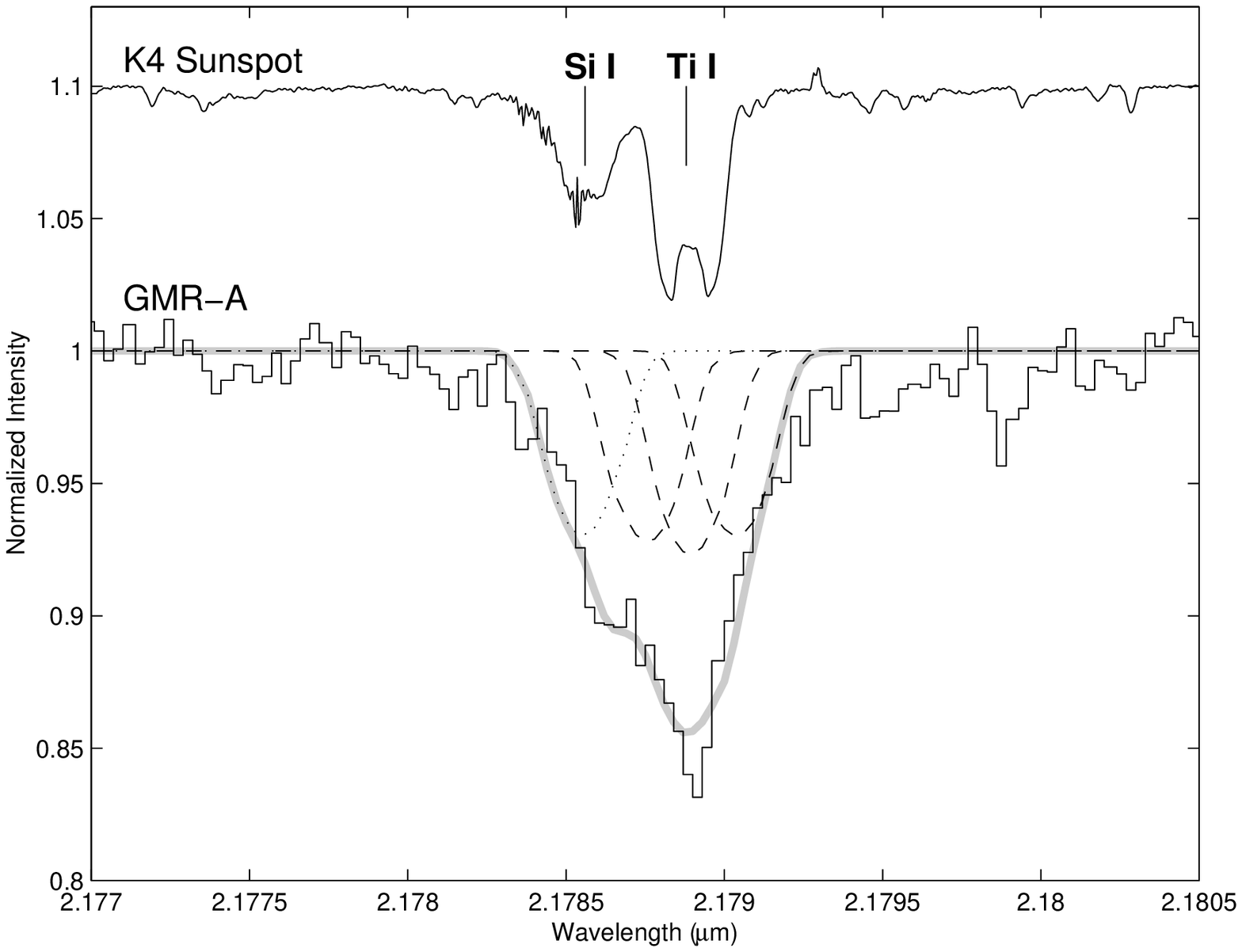} \figcaption[f7.eps]{Three
component Zeeman fit of the 2.17886 $\mu$m transition of
Ti~{\small I} in GMR-A. The three Zeeman components, broadened by the
instrumental profile and convolved with the rotational profile for
$v\,\sin i=23$ km~s$^{-1}$, are shown by dashed lines. A nearby
Si~{\small I} transition has been modeled in the same manner and
included in the fit (dotted line). The sum of all the components is
shown by the thick gray line, superimposed on the data.  A high
resolution spectrum of a sunspot with effective temperature similar to
a K4 star and a strongly split Ti~{\small I} line is shown for
comparison.  The splitting measured by the fit is $\Delta\lambda=0.14
\pm 0.07$ nm, corresponding to a magnetic field $3.8 \pm 2.0$ kG.
\label{fig:zeeman}}

\begin{deluxetable}{llll}
\tablecomments{Units of right ascension are hours, minutes, and seconds, and units of declination are degrees, arcminutes, and arcseconds.}
\tablecaption{Position of \object\ \label{tab:position}}
\tablehead{\colhead{Observation} & \colhead{Observing Band} & \colhead{RA} & \colhead{Dec.} \\
                            &                               & \colhead{(J2000)} & \colhead{(J2000)} }
\startdata
VLBA & 15 GHz & 05 35 11.802691 $\pm$ 0.000001 & -05 21 49.24660 $\pm$ 0.00004 \\
VLBA & 22 GHz & 05 35 11.802695 $\pm$ 0.000001 & -05 21 49.24660 $\pm$ 0.00003 \\
VLA & 15, 22, 43 GHz & 05 35 11.801 $\pm$ 0.001 & -05 21 49.27 $\pm$ 0.02 \\
BIMA & 86 GHz        & 05 35 11.80 $\pm$ 0.006 & -05 21 49.2 $\pm$ 0.1 \\
IR & 2.2 $\mu$m & 05 35 11.81 & -05 21 49.3\\
X-Ray & 1--- 10 keV & 05 35 11.8 & -05 21 49 \\
\enddata
\end{deluxetable}

\begin{deluxetable}{rr}
\tablecaption{BIMA Fluxes for \object\ \label{tab:bimafluxes}}
\tablecomments{Upper limits are $3 \sigma$ corrected for attenuation by the
primary beam.}
\tablehead{ \colhead{Day} & \colhead{$S_{86}$} \\
                                  & \colhead{(mJy)} } 
\startdata
 -8.29 & $<$    6   \\ 
  11.29 & $<$   12   \\ 
  14.29 & $<$   15   \\ 
  20.12 &   42 $ \pm $    9 \\ 
  20.17 &   39 $ \pm $    9 \\ 
  20.21 &   18 $ \pm $    7 \\ 
  20.25 &  122 $ \pm $    7 \\ 
  20.29 &  136 $ \pm $    7 \\ 
  20.33 &  146 $ \pm $    7 \\ 
  20.38 &  162 $ \pm $    8 \\ 
  23.23 & $<$   25   \\ 
  24.23 & $<$   13   \\ 
  30.38 & $<$   27   \\ 
  35.23 & $<$    6   \\ 
  38.16 &   11 $ \pm $    3 \\ 
\enddata
\end{deluxetable}

\begin{deluxetable}{rrrrrrrr}
\tablecaption{VLA Fluxes for \object\ \label{tab:vlafluxes}}
\tablecomments{$S_f$ and $V_f$ are the total intensity and circularly polarized flux density at frequency $f$.  Upper limits given for $V_f$ are $2\sigma$ limits on the absolute value.  UT days are given for the
year 2003.}
\tablehead{ \colhead{UT Day} & \colhead{$S_{8.4}$} & \colhead{$S_{15}$} & \colhead{$S_{22}$} & \colhead{$S_{43}$} & \colhead{$V_{15}$} & \colhead{$V_{22}$} & \colhead{$V_{43}$}\\
                                  & \colhead{(mJy)}   & \colhead{(mJy)}   & \colhead{(mJy)}   & \colhead{(mJy)}   & \colhead{(mJy)}   & \colhead{(mJy)}   & \colhead{(mJy)}  } 
\startdata
 22.23 & $<$   33   & \dots       &  85.7 $ \pm $   0.7 &  76.9 $ \pm $   1.0 & \dots       & $<$   1.4   & 
$<$   2.8   \\ 
 24.23 & \dots       &  18.5 $ \pm $   1.7 &  23.7 $ \pm $   0.6 &  16.6 $ \pm $   0.5 & $<$   2.6   &  -1.2 $ \pm $   0.3 & 
 -1.9 $ \pm $   0.5 \\ 
 27.26 & \dots       &  43.6 $ \pm $   1.6 &  43.7 $ \pm $   0.5 &  35.7 $ \pm $   0.4 & $<$   1.4   &  -2.2 $ \pm $   0.2 & 
 -3.4 $ \pm $   0.5 \\ 
 30.24 & \dots       &  32.4 $ \pm $   1.7 &  30.1 $ \pm $   0.8 &  26.6 $ \pm $   0.5 & $<$   2.0   & $<$   1.0   & 
$<$   2.3   \\ 
 31.25 & \dots       &  33.8 $ \pm $   1.6 &  34.7 $ \pm $   0.6 &  25.9 $ \pm $   0.5 & $<$   1.9   &  -0.6 $ \pm $   0.2 & 
$<$   1.9   \\ 
 32.17 & \dots       &  27.3 $ \pm $   1.6 &  19.2 $ \pm $   0.6 &  15.3 $ \pm $   0.4 & $<$   1.8   &  -1.1 $ \pm $   0.2 & 
$<$   2.1   \\ 
 33.21 & \dots       &  41.6 $ \pm $   1.6 &  46.8 $ \pm $   0.6 &  28.0 $ \pm $   0.5 &  -1.3 $ \pm $   0.3 &  -1.8 $ \pm $   0.2 & 
 -1.9 $ \pm $   0.4 \\ 
 34.21 & $<$   31   &  17.8 $ \pm $   1.0 &  19.9 $ \pm $   0.3 &  10.9 $ \pm $   0.3 &  -1.2 $ \pm $   0.2 &  -1.0 $ \pm $   0.1 & 
 -0.9 $ \pm $   0.3 \\ 
 36.23 & \dots       &  10.1 $ \pm $   2.5 &   8.7 $ \pm $   1.3 &   8.6 $ \pm $   0.5 & $<$   1.8   & $<$   3.1   & 
$<$   1.9   \\ 
 37.98 & \dots       &   9.9 $ \pm $   3.2 &  10.2 $ \pm $   1.1 &   5.5 $ \pm $   1.1 & $<$   4.0   & $<$   1.2   & 
$<$   1.6   \\ 
 42.13 & \dots       &  17.7 $ \pm $   2.1 &   7.1 $ \pm $   0.6 & $<$   1.0   & $<$   1.4   &  -0.6 $ \pm $   0.1 & 
$<$   1.3   \\ 
 45.04 & $<$   36   &   7.5 $ \pm $   2.1 & $<$   2.9   &   2.6 $ \pm $   0.5 & $<$   1.8   & $<$   1.8   & 
$<$   1.6   \\ 
 48.02 & $<$   33   &  27.4 $ \pm $   1.9 &   6.6 $ \pm $   0.7 & $<$   1.7   & $<$   3.5   & $<$   1.5   & 
$<$   1.3   \\ 
 53.17 & $<$   21   &  10.2 $ \pm $   1.6 &   9.4 $ \pm $   0.6 &   9.2 $ \pm $   0.3 &  -1.2 $ \pm $   0.2 & $<$   0.9   & 
 -1.0 $ \pm $   0.2 \\ 
 55.10 & \dots       &   7.3 $ \pm $   1.8 &   4.7 $ \pm $   0.7 &   6.7 $ \pm $   0.3 & $<$   1.5   & $<$   0.9   & 
$<$   1.2   \\ 
 59.12 & $<$   26   &  25.8 $ \pm $   1.9 &  10.2 $ \pm $   0.8 & $<$   1.5   &  -0.7 $ \pm $   0.2 &  -1.0 $ \pm $   0.1 & 
$<$   1.1   \\ 
 76.06 & $<$   18   &  24.9 $ \pm $   1.1 &  19.5 $ \pm $   0.5 &  15.9 $ \pm $   0.3 & $<$   1.2   &  -0.5 $ \pm $   0.1 & 
 -1.9 $ \pm $   0.3 \\ 
 83.04 & $<$   21   &  10.9 $ \pm $   1.0 &   9.1 $ \pm $   0.4 &   5.0 $ \pm $   0.2 &  -0.9 $ \pm $   0.1 &  -0.9 $ \pm $   0.1 & 
$<$   1.2   \\ 
 88.96 & $<$   36   &   8.8 $ \pm $   1.4 &   6.1 $ \pm $   0.5 &   3.2 $ \pm $   0.4 & $<$   1.0   &  -0.9 $ \pm $   0.1 & 
 -0.6 $ \pm $   0.1 \\ 
\enddata
\end{deluxetable}

\begin{deluxetable}{lrlrrrl}
\tablecaption{Infrared Photometry \label{tab:irphot}}
\tablehead{\colhead{Star} & \colhead{Date} & \colhead{Telescope} & \colhead{J} & \colhead{H} & \colhead{K$_S$} & \colhead{Reference} }
\tablecomments{Star S is the bright K band object 6$^{\prime\prime}$ South of \object.  Star W is a bright K
band object 6$^{\prime\prime}$ South and 20$^{\prime\prime}$ West of \object.  HC2000 refers to Hillenbrand \& Carpenter (2000). }
\startdata
\object\ & 02/09/99 & Keck/NIRC & \dots & 11.9 & 9.6 & HC2000 \#573 \\
\object\ & 01/22/03 & Keck/NIRC & 16.0 & 11.98 & 9.61 & This paper \\
\object\ & 01/24/03 & CTIO & 15.8 & 11.9 & 9.7 & This paper \\
\hline
Star W & 02/09/99 & Keck/NIRC & \dots & 10.5 & 9.9 & HC2000 \#554 \\
Star W & 01/22/03 & Keck/NIRC & 11.1 & 10.38 & 9.75 & This paper \\
Star W & 01/24/03 & CTIO & 11.2 & 10.3 & 9.8 & This paper \\
\hline
Star S & 02/09/99 & Keck/NIRC & \dots & 12.0 & 11.6 & HC2000 \#555 \\
Star S & 01/22/03 & Keck/NIRC & 12.7 & 12.1 & 11.5 & This paper \\
Star S & 01/24/03 & CTIO & 12.7 & 12.0 & 11.6 & This paper \\
\enddata
\end{deluxetable}

\end{document}